# High-fidelity spin readout via the double latching mechanism

Haruki Kiyama*[1,2], Danny van Hien[2], Arne Ludwig[3], Andreas D. Wieck[3], and Akira Oiwa[2,4,5,6]

[1]Graduate School and Faculty of Information Science and Electrical Engineering, Kyushu University, 744 Motooka, Nishi-ku, Fukuoka, 819-0395, Japan

[2]SANKEN, Osaka University, 8-1 Mihogaoka, Ibaraki, Osaka 567-0047, Japan

[3]Lehrstuhl für Angewandte Festkörperphysik, Ruhr-Universität Bochum, Universitätsstraße 150, Gebäude NB, D-44780 Bochum, Germany

[4]Center for Spintronics Research Network (CSRN), Graduate School of Engineering Science, Osaka University, 1-3 Machikaneyama, Toyonaka, Osaka 565-0871, Japan

[5]Center for Quantum Information and Quantum Biology, Osaka University, 1-3 Machikaneyama, Toyonaka, Osaka 560-0043, Japan

[6] Spintronics Research Network Division, Institute for Open Transdisciplinary Research (OTRI), Osaka University, Osaka 565-0871, Japan

*Corresponding author:  kiyama@ed.kyushu-u.ac.jp

**ABSTRACT**

Projective measurement of single electron spins, or spin readout, is among the most fundamental technologies for spin-based quantum information processing. Implementing spin readout with both high-fidelity and scalability is indispensable for developing fault-tolerant quantum computers in large-scale spin-qubit arrays. To achieve high fidelity, a latching mechanism is useful. However, the fidelity can be decreased by spin relaxation and charge state leakage, and the scalability is currently challenging. Here, we propose and demonstrate a double-latching high-fidelity spin readout scheme, which suppresses errors via an additional latching process. We experimentally show that the double-latching mechanism provides significantly higher fidelity than the conventional latching mechanism and estimate a potential spin readout fidelity of 99.94% using highly spin-dependent tunnel rates. Due to isolation from error-inducing processes, the double-latching mechanism combined with scalable charge readout is expected to be useful for large-scale spin-qubit arrays while maintaining high fidelity.

## Introduction

Projective measurement of qubits is a prerequisite for quantum information processing[1], and its implementation with high fidelity is needed for fault-tolerant quantum computation[2]. For spin qubits in semiconductor quantum dots (QDs), which are a promising platform for large-scale quantum computers[3], single-shot spin readout has been demonstrated using various spin-to-charge conversion processes and subsequent single-electron charge sensing[4-7]. To improve the spin readout fidelity, a latching mechanism, in which spin states are mapped to metastable charge states that have longer lifetimes and enhanced charge readout signals, has been developed[8-11]. This latching mechanism provides a readout fidelity greater than 99%[10,11] and has been utilized in silicon[12,13] and germanium spin-qubit systems[14]. However, errors still occur due to charge state leakage and spin relaxation in the conventional latching scheme[11], leaving room for further improvement.

In this work, we develop a new spin readout scheme using a double-latching (DL) mechanism. This scheme significantly improves the spin readout fidelity by using two consecutive gate-voltage pulses to enhance the charge signal and suppress spin relaxation and charge state leakage. We investigate the charge state leakage process in detail and optimize the DL spin readout to show the maximal fidelity. With the optimized spin readout, the spin readout fidelity is estimated to be as high as 99.94% from the observed spin-dependent tunnel rates, which is the highest ever reported for spin qubits and is well above the fault-tolerance threshold.

The DL spin readout will mitigate the requirement for device properties related to spin and charge relaxation rates. Moreover, for practical spin-based quantum computers, it is also important to implement spin readout in large-scale two-dimensional QD arrays[3, 15]. Conventional charge sensors are placed at the peripheries of arrays for direct coupling to electron reservoirs. Therefore, in large-scale QD arrays, many of the spin qubits are located far from the sensors, which reduces the fidelity of the charge readout. Gate-based spin readout is an alternative that is capable of measuring spins isolated from electron reservoirs[16-19]; however, the signal amplitude is not yet sufficient to achieve high-fidelity readout. Spin readout with a high charge sensitivity in large arrays may be achieved by spin shuttling[20,21], sequential SWAP gates[22] or cascaded spin-to-charge conversion[23], to bring electrons into the vicinity of the charge sensors, although errors remain due to relaxation processes and SWAP gate infidelity. However, the fidelity of these spin readout schemes is limited by the relaxation processes of the spin or charge during charge sensing. As we discuss later, the DL mechanism presented in this work should be useful for implementing high-fidelity spin readout in large-scale QD arrays by combining it with scalable charge readout, such as charge shuttling, charge cascades, and gate dispersive charge readout.

## Results
### Double-latching mechanism

The aim of the DL mechanism is ultimate suppression of relaxation or leakage by applying an additional voltage pulse immediately after the conventional latching mechanism is completed. The DL spin readout is implemented in a gate-defined double quantum dot (DQD) with a proximal charge sensor (Fig. 1a). An essential condition is that the QD–reservoir tunnel coupling is significantly different between the two QDs, as indicated by clear and blurred (or invisible) charge transition lines in Fig. 1b. More details about the device used in the experiments are provided in the next section.

The pulse sequence of the DL spin readout and corresponding energy level diagrams are depicted in Figs. 1b and 1c, respectively. The sequence consists of five stages: (i) initialization at I, (ii) Pauli spin blockade (PSB) at P, (iii) first latching at L, (iv) second latching, and (v) charge readout at M. The details of the pulse sequence used in the experiments presented below are provided in Supplementary Fig. 1 and Supplementary Note 1. First at I, the DQD is randomly initialized to either a spin-singlet or spin-triplet state by loading an electron from the left reservoir to QD2 via QD1 and waiting at I for mixing of (1,1) spin states. Then, the DQD is ramped to P to map the singlet and triplet states to the (2,0) and (1,1) charge states, respectively, by PSB[24], where ($N_1$, $N_2$) indicates the charge state of the DQD with $N_1$ and $N_2$ electrons in QD1 and QD2, respectively. This ramp is sufficiently slow to convert (1,1)S to (2,0)S adiabatically[25]. Next, the DQD is pulsed to L to perform the first latching, which converts (2,0) to (1,0) while maintaining (1,1) unchanged because of the large tunnel rate difference between the singlet and triplet states. The tunnel rates of the singlet and triplet states are denoted as $\varGamma_S^L$ and $\varGamma_T^L$, respectively, in Fig. 1c. This first latching enhances the contrast of the charge sensor signal and hence significantly improves the charge readout fidelity. Fidelity enhancement by the first latching has already been reported[11] and is referred to as the single-latching (SL) spin readout in this work.

Before describing the next stage of the DL spin readout sequence, we discuss the errors that occur during the first latching[11]. An inter-dot charge admixture causes unwanted transitions from (1,1) to (1,0) at L, hereinafter referred to as charge state leakage. In addition, cotunneling events may also cause unwanted transitions. These unwanted tunnelling events become more efficient when the interdot- and QD1–reservoir tunnel couplings become large. The fidelity of the SL spin readout is reduced by these error mechanisms, which can occur until the charge readout is completed in an integration time of typically a few microseconds[26].

After a waiting time in the first latching stage, the second latching pulse from L to M is applied to suppress the error due to spin relaxation and charge state leakage. At M, the interdot charge admixture is negligibly small, because the gate-voltage conditions are detuned far from the (2,0)–(1,1) resonance. Moreover, the lowest two-electron state is (1,1) at M. Therefore, the spin relaxation from (1,1)T to (1,1)S is not followed by a transition to (2,0)S, maintaining the correct spin–charge correspondence. With the second latching and charge sensing at M, the spin readout fidelity is expected to be significantly improved.

We numerically evaluated the effect of the DL mechanism by calculating the error due to spin relaxation at P and the errors of the latching processes in the SL and DL mechanisms. Figure 1d shows the latching errors as a function of the slower tunnel rate $\Gamma_{\text{Slow}}$ for two cases $\Gamma_S^L > \Gamma_T^L$ ($\Gamma_{\text{Slow}} = \Gamma_T^L$) and $\Gamma_T^L > \Gamma_S^L$ ($\Gamma_{\text{Slow}} = \Gamma_S^L$) [the details of the calculations are provided in Supplementary Note 2]. The latter case can be realized under gate voltage conditions that differ from those in Fig. 1b, as demonstrated later in Fig. 4. In the SL mechanism, the latching error is larger than the error caused by spin relaxation for any $\Gamma_S^L$ value when $\Gamma_S^L > \Gamma_T^L$ (red dashed curve). On the other hand, the error can be reduced when $\Gamma_T^L > \Gamma_S^L$ (blue dashed curve) by latching the triplet before the spin relaxation for $\Gamma_{\text{Slow}} < \Gamma_{\text{rel}}$. The DL mechanism further reduces the error for a wide range of $\Gamma_{\text{Slow}}$ values (red and blue solid curves). The error reduction is not significant when the error is already small in the SL mechanism. The DL mechanism greatly improves the SL readout fidelity while only requiring the application of additional gate voltage pulses, particularly in devices with fast spin relaxation and charge leakage, which offers a high-fidelity spin readout scheme with broad applicability.

**Device setting**

The DL spin readout was experimentally demonstrated in a DQD with a QD charge sensor (Fig. 1a) formed in a silicon-doped GaAs/AlGaAs heterostructure with a two-dimensional electron gas located 100 nm below the surface. Two plunger gates, P1 and P2, were used to control the electrochemical potentials of QD1 and QD2, respectively. Three barrier gates, B1, B2 and B3, were used to control the tunnel couplings between the two dots or between a dot and the reservoir. The device was cooled in a dilution refrigerator to a base temperature of 20 mK and an electron temperature of 160 mK without an external magnetic field, unless otherwise noted.

In the DL spin readout experiments, a large QD1–reservoir tunnel coupling (~120 MHz), a small QD2–reservoir tunnel coupling (~400 Hz), and an interdot tunnel coupling of 1.48 GHz were established. The singlet–triplet energy splitting was evaluated as $\Delta_{ST}$ ~ 250 μeV. The details for the evaluation of $\Delta_{ST}$ and dot–reservoir and interdot tunnel couplings are provided in Supplementary Fig. 4 and Supplementary Notes 3 and 4, respectively. The sensor dot was used to measure the charge configuration of the DQD by radio-frequency reflectometry.

Figure 1b shows a charge stability diagram of the DQD as a function of the gate voltages $\Delta V_{P1}$ on P1 and $\Delta V_{P2}$ on P2. The DQD was set around the (2,0)–(1,1) transition for the spin readout experiments. The boundary of the ground charge state of QD2 (represented by the grey solid lines in Fig. 1b) was not observed because $\Delta V_{P2}$ was swept faster than the QD2–reservoir tunnel rate of < 1 kHz. Instead, the QD2 charge transition from (1,0) to (1,1) occurred at the (1,0)–(2,0) resonance by electron loading to QD1 to form (2,0) and subsequent interdot tunnelling to (1,1). The transition from (2,0) to (2,1) was visible due to cotunneling.

## Enhancement of the charge sensor signal and decay time

First, we experimentally confirmed that the charge sensor signal is enhanced compared to that in the conventional PSB spin readout due to the latching mechanisms. Figure 2a shows a histogram of the charge sensor signals measured at P in the PSB spin readout with an integration time of 3 μs. The histogram was fitted well by two Gaussian distributions with spin relaxation during the readout considered[6]. The signal-to-noise ratio was as low as ~2.5. Figures 2b and 2c show the sensor signal histograms measured at L and M by the SL spin readout and DL spin readout, respectively, with an integration time of 1 μs. In both results, the signal-to-noise ratio was significantly improved to ~10 despite the shorter integration time compared to the PSB readout. This signal enhancement arises because the total electron number changes by one. The low signal count for (1,1) in the SL spin readout (Fig. 2b) is attributed to the relatively high $\Gamma_T^L$ due to interdot charge admixture.

Figure 2d shows the charge sensor signal as a function of time at P, averaged over 100,000 single-shot PSB spin readouts. Fitting with a single-exponential function yielded a triplet-to-singlet spin relaxation rate of $\Gamma_{rel}$ = 55.5 ± 0.8 kHz. In the SL spin readout, the sensor signal decayed more rapidly at L, as shown in Fig. 2e, which is attributed to charge state leakage due to the admixing of (1,1)T and (2,0)T. The DL spin readout drastically extended the decay time, as shown in Fig. 2f, by suppressing both the charge admixture and the spin relaxation. Figures 2e and 2f show that the triplet tunnel rate $\Gamma_T^L$ was 740 ± 20 kHz at L and 0.358 ± 0.005 kHz at M, respectively.

To verify that the DL mechanism can be used for spin readout, the charge sensor signal at M was measured as a function of the waiting time at P, as shown in Fig. 2g. The signal decay rate was 58 ± 2 kHz, which is comparable to the value of $\varGamma_{\text{rel}}$ (Fig. 2d). This result indicates that the DL spin readout is useful for distinguishing between singlet and triplet spin states.

**Readout fidelity and optimization**

The fidelity of the PSB spin readout was 79.5% according to the single-shot histogram (Fig. 2a) (details are provided in Supplementary Notes 5). Spin relaxation during an arming time of 3 μs, which is the waiting time due to the measurement bandwidth, and a charge readout integration time of 3 μs was considered. The fidelity also includes the PSB error due to charge nonadiabaticity. This error was $7.0 \times 10^{-8}$ according to the Landau-Zener formula, $\exp\left[-2\pi(\sqrt{2}t_c)^2 \Delta t/\hbar\Delta E\right]$, where $t_c$ = 1.48 GHz is the interdot tunnel coupling, $\Delta t$ = 20 ns is the ramp time from I to P, and $\Delta E \approx 870$ μeV is the change in the detuning from I to P.

Regarding the fidelity of the SL spin readout and DL spin readout, we considered the detuning dependence of the tunnel rates because the fidelity highly depended on the tunnel rates, as shown in Fig. 1d. Figure 3a shows the sensor signal measured by the DL spin readout as a function of the waiting time at L. The signal shows fast and slow decays corresponding to tunnelling from the singlet and triplet states, respectively, to (1,0). The tunnel rates were evaluated by fitting the data with a double-exponential function and are plotted in Fig. 3b. For the singlet state, the tunnel rate $\varGamma_S^L$ varied from low (~1 MHz) at positive detuning values to high (~120 MHz) at negative detuning values, consistent with the charge state transition from (1,1) to (2,0). The triplet tunnel rate $\varGamma_T^L$ showed a slight increase with decreasing detuning. This result indicates that the charge configuration of the triplet state was predominantly (1,1) in the measured detuning range, consistent with $\varDelta_{\text{ST}} \sim 250$ μeV obtained via excited state spectroscopy (Supplementary Fig. 4). The model fit provided a smaller value of $\varDelta_{\text{ST}} = 197\pm7$ μeV (details are provided in Supplementary Note 7), presumably due to weaker confinement in the gate-voltage condition for the spin readout experiments.

The fidelity of the SL spin readout was calculated from the single-shot histogram (Fig. 2b), along with the tunnel rates (Fig. 3b) and the spin relaxation rate (Fig. 2d) (details are provided in Supplementary Note 6). The calculated fidelity is shown by grey open circles in Fig. 3c as a function of the detuning, in which spin relaxation and charge state leakage during the arming time were taken into account but the detuning dependence of the spin relaxation rate was ignored[27]. Relaxation of the triplet state to (1,0) during the arming time significantly decreased the fidelity, as shown by the grey curves in Fig. 3c for arming times of 0.1 μs and 1 μs. In the experiment with an arming time of 1 μs, the maximum fidelity was as low as 68.7% for a detuning of 63 μeV. This accounted for the small peak height for the (1,1) state in Fig. 2b. The significant decrease in the SL readout fidelity with increasing arming time arises because $\Gamma_T^L$ is much greater than $\Gamma_{rel}$ due to fast charge leakage. Suppression of the charge leakage requires either a large $\Delta_{ST}$, a low QD1–reservoir coupling or a low interdot tunnel coupling. The realization of a large $\Delta_{ST}$ imposes more restrictions on the materials and structures of QD devices. A low QD1–reservoir coupling causes slow initial latching, which increases the spin relaxation error. A low interdot tunnel coupling causes more error in the spin-to-charge conversion by PSB. For these reasons, implementation of high-fidelity SL readout is possible only in devices with excellent properties.

In the DL spin readout, the readout fidelity consists of three factors: (i) the PSB fidelity, (ii) the fidelity of the DL mechanism, hereinafter referred to as the latching fidelity, which includes errors due to spin relaxation and charge state leakage during the arming time, and (iii) the charge readout fidelity. The latching fidelity was calculated for each of the singlet and triplet states using the observed tunnel rates and spin relaxation rate (the details of the calculations are provided in Supplementary Note 2) and was optimized in our experiment by tuning the detuning value of L and the waiting time there[28]. The black solid circles in Fig. 3c show the overall DL spin readout fidelity, averaged over the singlet and triplet states, at the optimal waiting time (Fig. 3d). The relaxation error at M during the arming time of 1 μs is $3.7 \times 10^{-4}$, and the charge readout error at M with an integration time of 1 μs is $7.9 \times 10^{-5}$. These errors are relatively small, and the overall readout fidelity was dominated by the latching fidelity. Because of this error suppression, the DL spin readout fidelity was significantly higher than the SL spin readout fidelity. The DL spin readout fidelity reached a maximum value of 97.13% at a detuning value of −20 μeV, at which the tunnel rate ratio of the singlet state to the triplet state, $\Gamma_S^L/\Gamma_T^L$, also reached a maximum value of ~100.

**High-fidelity readout**

We now demonstrate high-fidelity DL spin readout under a different device setting. The QD1–reservoir tunnel coupling was set to ~200 Hz, which was lower than the QD2–reservoir tunnel coupling of ~2 GHz. The interdot tunnel coupling and $\Delta_{ST}$ were 930 MHz and ~ 400 μeV, respectively. Accordingly, the gate voltage pulse sequence was modified to convert (1,1)T to (1,0) while keeping (2,0)S unchanged (Fig. 4a and Supplementary Fig. 5), which more efficiently suppressed a spin relaxation error because the (1,1)T excited state was quickly converted to the (1,0) charge state. Moreover, the difference in the QD1–reservoir and QD2–reservoir tunnel couplings was enhanced by a factor of ~80 compared to that under the condition for Figs. 1-3. This condition, together with the smaller interdot tunnel coupling and the larger $\Delta_{ST}$, is expected to contribute to a much larger difference between $\Gamma_S^L$ and $\Gamma_T^L$, which will improve the latching fidelity.

The probability of detecting a singlet state via the DL spin readout is shown in Fig. 4b as a function of the waiting time at L. A double-exponential decay was observed, as in the previous condition (Fig. 3a). The time constants of the fast and slow decays were $\Gamma_T^L$ = 2.0±0.1 GHz and $\Gamma_S^L$ = 109±5 kHz, respectively; hence, $\Gamma_T^L/\Gamma_S^L \sim 1.8 \times 10^4$. The detuning value at L for this measurement was chosen such that the fast decay component had sufficient data points, whereas the slow decay component had the lowest tunnel rate.

As in the previous condition, the latching fidelity in the DL mechanism was calculated using the $\Gamma_T^L$ and $\Gamma_S^L$ observed in Fig. 4b. We used the same $\Gamma_{rel}$ as in the previous condition (Fig. 2d), ignoring its gate-voltage dependence. Figure 4c shows the latching fidelity as a function of the waiting time at L. The waiting time was optimized to 5.0 ns to obtain the highest average latching fidelity of 99.97%. The PSB error due to charge nonadiabaticity was estimated to be $2.6 \times 10^{-6}$ with $t_c$ = 930 MHz and $\Delta E \approx 440$ μeV. The error in the charge readout at M was estimated to be $1.1 \times 10^{-4}$ from the charge sensor signal distribution (not shown). The charge relaxation error at M during the arming time of 1 μs was $2.2 \times 10^{-4}$ from the relaxation rate of 216 Hz. By considering these errors, the overall fidelity of the DL spin readout was then evaluated to be 99.94%. Although several possible errors, such as excitation and relaxation during the PSB ramp from I to P, were not considered here, the estimated spin readout fidelity was comparable to the highest value ever reported for semiconductor spin qubits[23]. The readout fidelity, including any errors, will be evaluated via experiments with high-fidelity state preparation[29]. Although we tuned the gate voltage to increase $\Delta_{ST}$ here, the high-fidelity DL spin readout does not require either a long spin relaxation time or an extremely large $\Delta_{ST}$. Since tuning asymmetric tunnel couplings for the DL mechanism is an inherent ability of gate-defined QD systems, high-fidelity DL spin readout will be feasible for a variety of QD devices.

## Discussion

The DL spin readout demonstrated in a DQD in this work will also be useful in larger QD arrays. In the case of a triple QD (Fig. 5a), PSB is first performed in the left and middle

QDs (labelled as QD$_i$ and QD$_{i+1}$, respectively, in Fig. 5a). An electron resides in each QD$_i$ and QD$_{i+1}$ for a spin-triplet state, whereas two electrons reside in QD$_i$ for a spin-singlet state. The first latching is then implemented by lowering the electrochemical potential of the right QD (QD$_{i+2}$). The electron in QD$_{i+1}$ tunnels to QD$_{i+2}$ instead of the reservoir, whereas the electrons in QD$_i$ remain there. Finally, the second latching is performed by increasing the electrochemical potential of QD$_{i+1}$ to suppress the charge state leakage from QD$_i$ to QD$_{i+2}$ due to the charge admixture.

The abovementioned DL mechanism in a triple QD may be extended to larger arrays, including two-dimensional QD arrays. In arrays partially filled with spin qubits[15] (Fig. 5b), after the DL mechanism, the electron in QD$_{i+2}$ can be sent to the vicinity of the charge sensor by charge shuttling for a triplet state, whereas no electron is sent for a singlet state. The small QD–reservoir tunnel coupling needed for the DL readout in a DQD is substituted by Coulomb blockade of electron tunneling from QD$_i$ to neighbouring QDs other than QD$_{i+1}$. The DL mechanism can also be combined with either electron cascades or gate-based charge readout (Fig. 5b) instead of the charge shuttling, in which the QD arrays may be more densely filled with spin qubits than in the charge-shuttling case. Owing to the isolation from error-inducing processes by the DL mechanism, the spin readout fidelity is expected to be as high as that in the case of a triple QD, independent of the array size. Moreover, the fidelity of the gate-based charge readout is rather low compared to the radio-frequency reflectometry of the proximal charge sensor. The DL mechanism allows the integration time to be increased to improve the charge readout fidelity with a negligible increase of relaxation errors.

In conclusion, we have proposed a spin readout scheme with high fidelity and broad applicability using the DL mechanism, which significantly improves the readout fidelity by suppressing spin and charge relaxation and enhancing the charge readout signal. In a GaAs DQD with fast spin and/or charge relaxation, we experimentally demonstrated that the DL spin readout showed much higher fidelity than the conventional SL readout. Furthermore, our experimental demonstration implied an optimized readout fidelity greater than 99.9%, comparable to the highest value ever reported for spin qubits. We discussed how the DL spin readout will be useful for high-fidelity spin readout in large-scale two-dimensional QD arrays. Moreover, although our demonstration used a GaAs DQD, the DL spin readout is applicable to silicon and germanium QD devices, in which high-fidelity multiqubit operations have been reported[14, 30-32]. Therefore, the DL spin readout is expected to advance the study of spin manipulation and spin dynamics in large-scale QD arrays and will contribute to the realization of fault-tolerant spin-based quantum computers in the future.

## Methods
### Device and setup
The device used in this work is a gate-defined DQD with an adjacent QD charge sensor formed in a GaAs/AlGaAs heterostructure. A two-dimensional electron gas is located 100 nm below the surface and has a carrier density of $1.2 \times 10^{11}$ cm$^{-2}$ and a mobility of $8.1 \times 10^5$ cm$^2$/Vs at 4.2 K. Surface Schottky gate electrodes were fabricated by electron-beam lithography and electron-beam evaporation of Ti and Au. Although the gate pattern was designed to define a quadruple QD and two sensing dots, some of the gates was used to form a DQD and a sensing dot.

The measurements were performed in a BlueFors LD dilution refrigerator at a base temperature of 20 mK and an electron temperature of 160 mK without an external magnetic field, unless otherwise noted. Voltage pulses were generated with a Tektronix AWG5014 arbitrary waveform generator and were applied to Gates P1 and P2 with DC voltages through bias-tees ($R = 5$ MΩ, $C = 100$ nF). Charge sensing was performed by radio-frequency reflectometry at a carrier frequency of 201.0 MHz. The reflected signal was amplified with a Cosmic Microwave Technologies CITLF3 cryogenic amplifier at 4 K, demodulated to the baseband at room temperature, low-pass filtered at a cut-off frequency of 300 kHz (Fig. 2a) or 1 MHz (other figures), and then sampled by a Spectrum M2p digitizer card at a sampling rate of 1 MS/sec.

## Data availability
The data that support the findings of this study are available from the corresponding author upon reasonable request.

## Acknowledgements
The authors thank Y. Kanai, D. Chiba, and K. Matsumoto for access to electron-beam lithography. This work was supported by JSPS KAKENHI (Grant No. 23H01793, 23H05458, 23K17764), the Casio Science Promotion Foundation, the Murata Science Foundation, the Support Center for Advanced Telecommunications Technology Research, the Okawa Foundation for Information and Telecommunications, the Mazda Foundation, the Cooperative Research Program of "Network Joint Research Center for Materials and Devices (MEXT)", JST (Moonshot R&D) (Grant No. JPMJMS2066), NRC Challenge

Program (QSP013), the Dynamic Alliance for Open Innovation Bridging Human Environment and Materials, and DFG via ML4Q (Grant No. 390534769).

## Author contributions

H. K. conceived, designed, and performed the experiments, and analyzed the data with inputs from A.O. A.L. and A.D.W. grew the heterostructure. H. K. and D. v. H fabricated the device. H. K wrote the manuscript with comments from all authors.

## Competing interests

The authors declare no competing interests.

## References

1. DiVincenzo, D. P. The Physical Implementation of Quantum Computation. *Fortschr. Phys.* **48**, 771 (2000).
2. Fowler, A. G., Stephens, A. M. & Groszkowski, P. High-threshold universal quantum computation on the surface code. *Phys. Rev. A* **80**, 052312 (2009).
3. Vandersypen, L. M. K. et al. Interfacing spin qubits in quantum dots and donors-hot, dense, and coherent. *npj Quantum Inf.* **3**, 1–10 (2017).
4. Elzerman, J. M. et al. Single-shot read-out of an individual electron spin in a quantum dot. *Nature* **430**, 431–435 (2004).
5. Hanson, R. et al. Single-shot readout of electron spin states in a quantum dot using spin-dependent tunnel rates. *Phys. Rev. Lett.* **94**, 196802 (2005).
6. Barthel, C., Reilly, D. J., Marcus, C. M., Hanson, M. P. & Gossard, A. C. Rapid single-shot measurement of a singlet-triplet qubit. *Phys. Rev. Lett.* **103**, 160503 (2009).
7. Kiyama, H., Nakajima, T., Teraoka, S., Oiwa, A. & Tarucha, S. Single-shot ternary readout of two-electron spin states in a quantum dot using spin filtering by quantum Hall edge states. *Phys. Rev. Lett*. **117**, 236802 (2016).
8. Studenikin, S. A. et al. Enhanced charge detection of spin qubit readout via an intermediate state. *Appl. Phys. Lett.* **101**, 233101 (2012).
9. Mason, J. D. et al. Role of Metastable Charge States in a Quantum-Dot Spin-Qubit Readout, *Phys. Rev. B* **92**, 125434 (2015).
10. Nakajima, T. et al. Robust single-shot spin measurement with 99.5% fidelity in a

quantum dot array. *Phys. Rev. Lett.* **119**, 017701 (2017).

11. Harvey-Collard, P. et al. High-fidelity single-shot readout for a spin qubit via an enhanced latching mechanism. *Phys. Rev. X* **8**, 021046 (2018).

12. Fogarty, M. A. et al. Integrated silicon qubit platform with single-spin addressability, exchange control and single-shot singlet-triplet readout. *Nat. Commun.* **9**, 4370 (2018).

13. Zhao, R. et al. Single-spin qubits in isotopically enriched silicon at low magnetic field. *Nat. Commun.* **10**, 5500 (2019).

14. Hendrickx, N. W. et al. A four-qubit germanium quantum processor. *Nature* **591**, 580 (2021).

15. Li, R. et al. A crossbar network for silicon quantum dot qubits. *Sci. Adv.* 4, eaar3960 (2018).

16. West, A. et al. Gate-based single-shot readout of spins in silicon. *Nat. Nanotechnol.* **14**, 437–441 (2019).

17. Urdampilleta, M. et al. Gate-based high fidelity spin readout in a CMOS device. *Nat. Nanotechnol*. **14**, 737–742 (2019).

18. Zheng, G. et al. Rapid gate-based spin read-out in silicon using an on-chip resonator. *Nat. Nanotechnol.* **14**, 742–746 (2019).

19. Borjans, F., Mi, X., & Petta, J. R. Spin Digitizer for High-Fidelity Readout of a Cavity-Coupled Silicon Triple Quantum Dot. *Phys. Rev. Appl.* **15**, 044052 (2021).

20. Baart, T. A. et al. Single-spin CCD. *Nat. Nanotechnol.* **11**, 330–334 (2016).

21. Yoneda, J. et al. Coherent spin qubit transport in silicon. *Nat. Commun.* **12**, 4114 (2021).

22. Sigillito, A. J., Gullans, M. J., Edge, L. F., Borselli, M. & Petta, J. R. Coherent transfer of quantum information in a silicon double quantum dot using resonant SWAP gates. *npj Quantum Inf.* **5**, 1–7 (2019).

23. van Diepen, C. J. et al. Electron cascade for distant spin readout. *Nat Commun*. **12**, 77 (2021).

24. Ono, K., Austing, D. G., Tokura, Y. & Tarucha, S. Current rectification by Pauli exclusion in a weakly coupled double quantum dot system. *Science* **297**, 1313-1317 (2002).

25. Petta, J. R. et al. Coherent Manipulation of Coupled Electron Spins in Semiconductor Quantum Dots. *Science* **309**, 2180–2184 (2005).

26. Barthel, C. et al. Fast sensing of double-dot charge arrangement and spin state with a radio-frequency sensor quantum dot. *Phys. Rev. B* **81**, 161308(R) (2010).

27. Barthel, C. et al. Relaxation and readout visibility of a singlet-triplet qubit in an

Overhauser field gradient. *Phys. Rev. B* **85**, 035306 (2012).

28. Osika, E. N. et al. Shelving and latching spin readout in atom qubits in silicon. *Phys. Rev. B* **106**, 075418 (2022).

29. Kobayashi, T. *et al.* Feedback-based active reset of a spin qubit in silicon. *npj Quantum Inf* **9**, 52 (2023).

30. Noiri, A. et al. Fast universal quantum gate above the fault-tolerance threshold in silicon. *Nature* **601**, 338–342 (2022).

31. Philips, S. G. J. et al. Universal control of a six-qubit quantum processor in silicon. *Nature* **609**, 919–924 (2022).

32. Takeda, K., Noiri, A., Nakajima, T., Kobayashi, T. & Tarucha, S. Quantum error correction with silicon spin qubits. *Nature* **608**, 682–686 (2022).

33. DiCarlo, L. et al. Differential charge sensing and charge delocalization in a tunable double quantum dot. *Phys. Rev. Lett.* **92**, 226801 (2004).

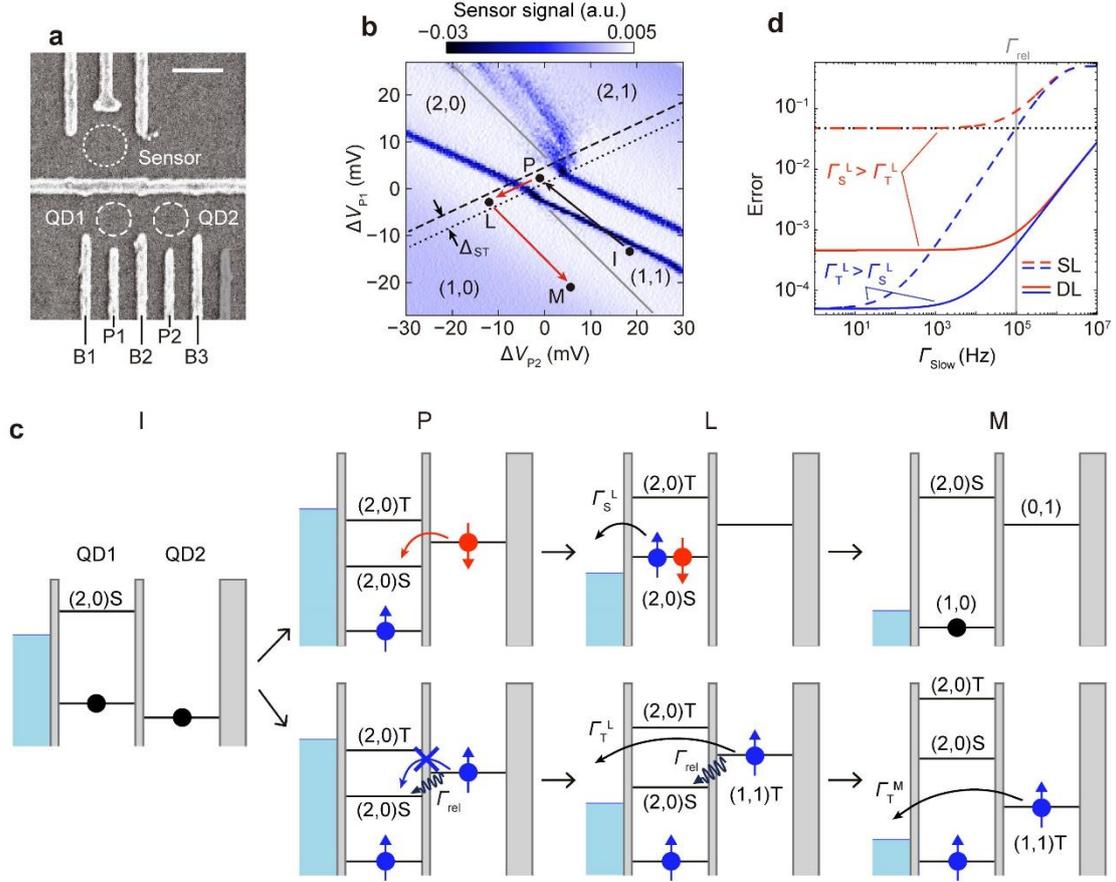

**Fig. 1 Device and double latching mechanism. a** Scanning electron micrograph of a device nominally identical to the one used in the experiments. The dashed circles and the dotted circle indicate a DQD and a sensor dot, respectively. An unused gate on the right of the DQD is coloured in grey. The scale bar corresponds to 200 nm. **b** Charge stability diagram of the DQD around the (2,0)–(1,1) transition as a function of the gate voltages $\Delta V_{P1}$ and $\Delta V_{P2}$. ($N_1$, $N_2$) indicates the ground state charge configuration of the DQD with $N_1$ and $N_2$ electrons in QD1 and QD2, respectively. $\Delta V_{P2}$ was increased at a rate of 100 V/s, while $\Delta V_{P1}$ was increased at a rate of less than 1 mV/s. The gate voltage conditions of the initialization (I), Pauli spin blockade (P), first latching (L), second latching and charge readout (M) are indicated by black circles. The dotted (dashed) line indicates the resonance between (2,0) and (1,1) for the singlet (triplet) state, estimated from the measured singlet–triplet splitting of $\Delta_{ST} = 250$ μeV. **c** Schematics of the DQD energy levels at I, P, L and M for singlet (top row) and triplet (bottom row) states. The curved arrows represent single-electron tunnelling processes. The wavy arrows indicate spin relaxation from a triplet state to a singlet state at a rate $\Gamma_{rel}$. **d** Latching errors in the SL (dashed lines) and DL (solid lines) mechanisms calculated for two cases, $\Gamma_S^L > \Gamma_T^L$ (red) and $\Gamma_T^L > \Gamma_S^L$ (blue), as a function of the slower tunnel rate, $\Gamma_{Slow}$. The faster tunnel rate

and the spin relaxation rate are assumed to be 100 MHz and 100 kHz (grey vertical line), respectively. The error due to spin relaxation at P within a 1 µs integration time is shown by a dotted line.

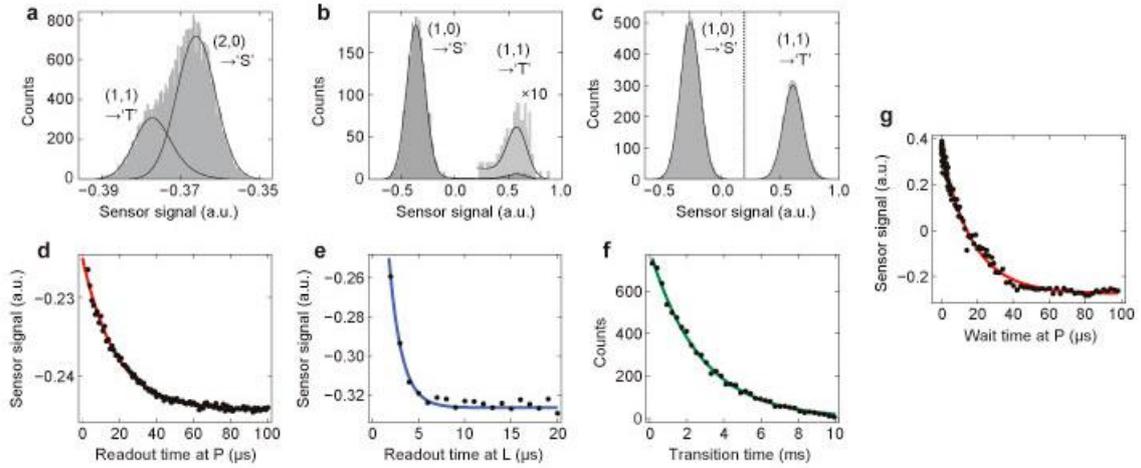

**Fig. 2 Single-shot readout statistics.** Histograms of single-shot charge sensor signals for **a** PSB spin readout, **b** SL spin readout and **c** DL spin readout. The black solid curves are the fits for two Gaussian distributions with spin relaxation. The dotted vertical line in **c** represents the threshold signal level to distinguish between the singlet and triplet states in the DL spin readout. **d** Charge sensor signal as a function of time at readout point P in the PSB spin readout. The signal was averaged over 100,000 single-shot measurements. **e** Same as **d** but measured at readout point L in the SL spin readout. **f** Distribution of the transition time from (1,1) to (1,0) at readout point M in the DL spin readout. **g** Charge sensor signal as a function of the waiting time at P measured in the DL spin readout. The signal was averaged over 2,000 single-shot measurements in **e** and **g**. The sensor signal values in **d**, **e** and **g** differed from those in **a**, **b** and **c**, respectively, due to changes in the charge sensor conditions.

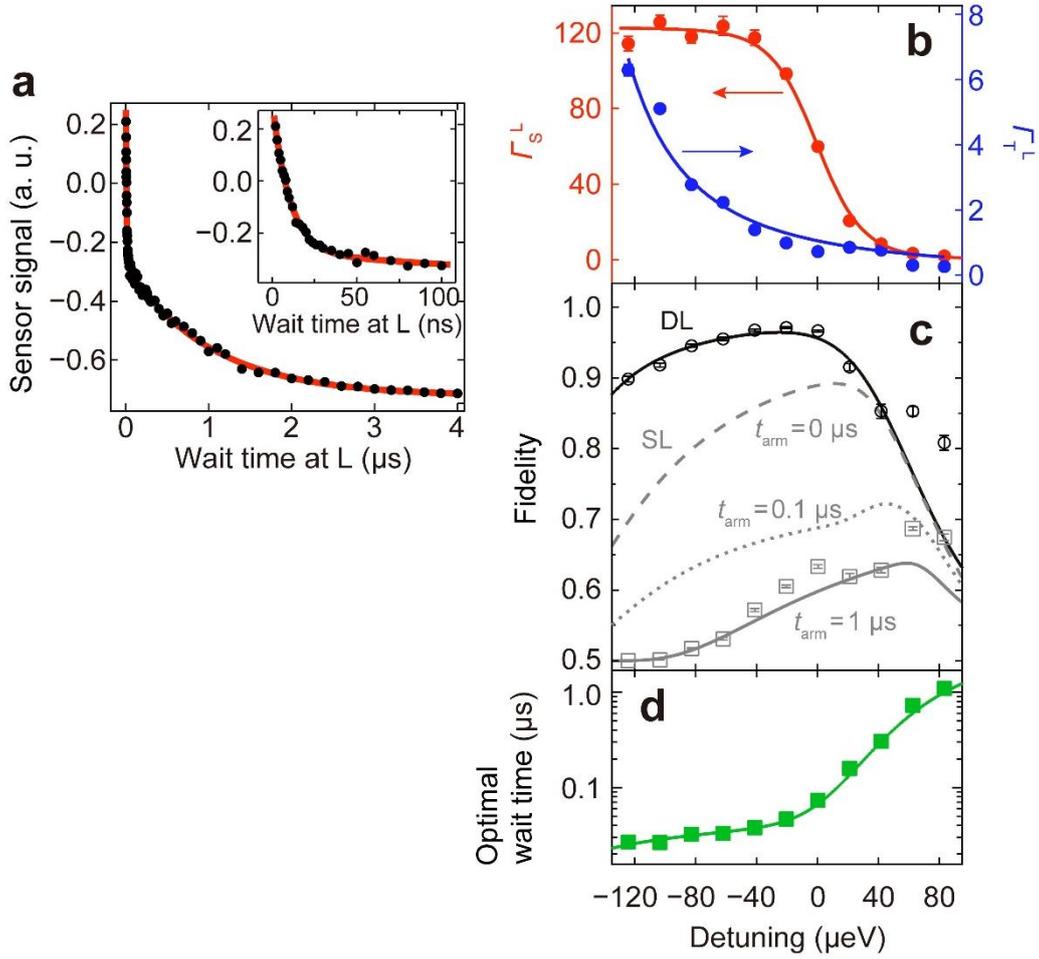

**Fig. 3 Optimization of the first latching condition. a** Charge sensor signal as a function of the waiting time at L in the DL spin readout. The inset shows an enlarged figure for a waiting-time range shorter than 100 ns. The red curve is the fit to a double exponential function. **b** Detuning dependence of the tunnel rates of the singlet (red) and triplet (blue) states at L, $\Gamma_S^L$ and $\Gamma_T^L$. The solid curves show fits to the model that considers tunnelling from hybridized charge states. **c** Fidelity of the DL spin readout (solid black circles) with an optimized waiting time between the first and second latching pulses shown in **d**. The black solid curve was calculated using the fitting results of the tunnel rates in **b**. The grey data show the SL spin readout fidelity calculated for an integration time of 1 μs and arming times of 0 μs (dashed curve), 0.1 μs (dotted curve), and 1 μs (open circles and solid curve).

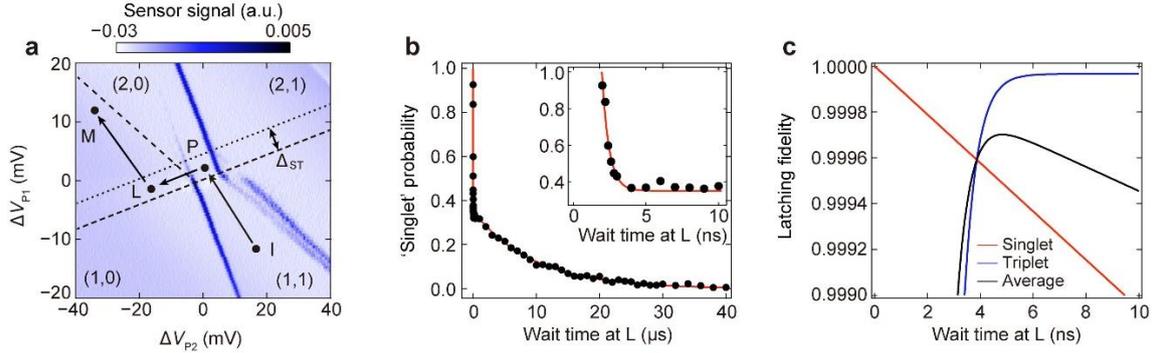

**Fig. 4 High-fidelity double-latching spin readout. a** Same as Fig. 1b but for the high-fidelity readout experiment with $\Delta_{ST} = 400$ μeV. **b** Probability of detecting a singlet as a function of the waiting time at L. The inset shows an enlarged figure for a waiting-time range shorter than 10 ns. The waiting time is offset by ~2 ns due to the rise time of the gate voltage pulse. The red curve is the fit to a double exponential function. **c** Latching fidelities calculated for the singlet state (red), the triplet state (blue) and their average (black) as a function of the waiting time at L. The tunnel rates used for the calculation are obtained from **b**. The waiting time offset due to the pulse rise time is ignored.

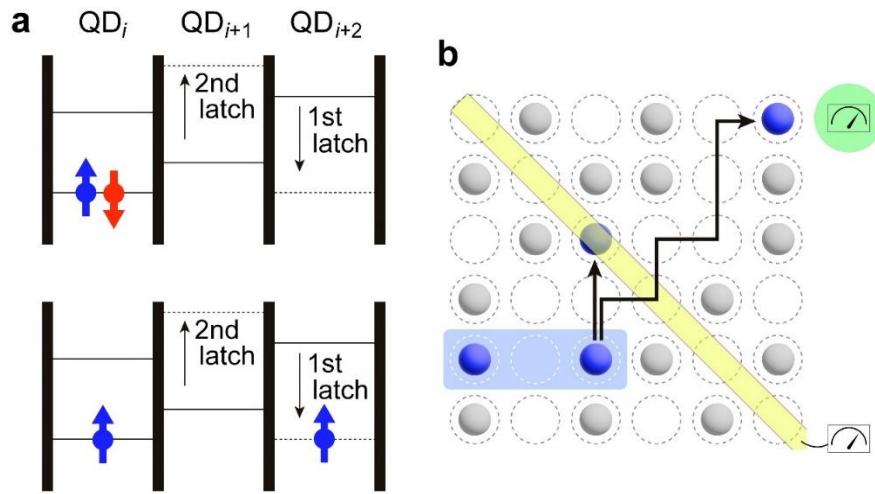

**Fig. 5 Double-latching spin readout in a 2D QD array. a** DL spin readout in a triple QD. Two electrons are in the left QD (QD$_i$) for the spin-singlet state (top), whereas the electrons are separated in distant QDs (QD$_i$ and QD$_{i+2}$) for the spin-triplet state (bottom). **b** Charge shuttling to the vicinity of the sensor (green circle) at the periphery of the QD array or to the gate electrode (yellow bar) for dispersive charge readout. The triple QD in **a** is shaded blue. The paths of the charge shuttling are shown by arrows. The electrons involved in the readout are shown by blue circles, and the other electrons are shown by grey circles.

17